\documentclass[letterpaper, 10 pt, conference]{ieeeconf}

\usepackage{color}
\usepackage{amsmath}
\usepackage{amsfonts}
\usepackage{amssymb}
\usepackage{bm}
\usepackage[cal=cm]{mathalfa}
\usepackage{pgfplots}
\usepackage{multirow}
\usepackage{mdframed}
\usepackage{cite}

\def\url#1{}

\DeclareMathOperator{\erf}{erf}

\title{\LARGE \bf Uncertainty-based Arbitration of Human-Machine Shared Control}
\author{Parker Owan, Joseph Garbini, and Santosh Devasia}

\begin{document}

\maketitle
\thispagestyle{empty}
\pagestyle{empty}

\begin{abstract}
Manufacturing requires consistent production rate and task success for sustainable operation.  Some manufacturing tasks require a semi-autonomous approach, exploiting the combination of human adaptability and machine precision and speed, to be cost effective.  The main contribution of this  paper is a new approach to determine the level of autonomy for human-machine shared control based on the automation uncertainty.   
Moreover, the haptic feedback is scaled by the level of autonomy to indicate machine confidence to the operator. 
Experimentation results, with a human-robot peg-in-a-hole testbed, show more than  5 times improvement in the error tolerance for task completion with the shared control approach when compared to a purely autonomous method. 
\end{abstract}

\section{Introduction}
Fully autonomous systems are becoming increasingly prevalent in manufacturing facilities throughout the world, some of which operate with minimal human supervision~\cite{Null2003}.  Some manufacturing operations such as aircraft production have been unable to adopt a fully autonomous policy because, according to Felder~\cite{Felder2011}: (i)~new aircraft design tends to be a modification of previous designs where autonomous production was not a primary design consideration; (ii)~the deliverable product is larger than most of the machines used in its assembly; and (iii)~the delivery rate is much lower than required rates of facilities that have transitioned to fully-autonomous production. In order to transition from the current state to increased autonomy in aircraft manufacturing, there is interest in semi-autonomous solutions wherein the human shares some aspect of control with the machine.  
Such semi-autonomous approach can allow for more cost-effective solutions to support aircraft manufacturing  when compared to the fully autonomous approach.  For such semi-autonomous manufacturing,  there is a need to develop shared control strategies wherein both the machine and the human are simultaneously managing a task.

\vspace{0.1in}
\par
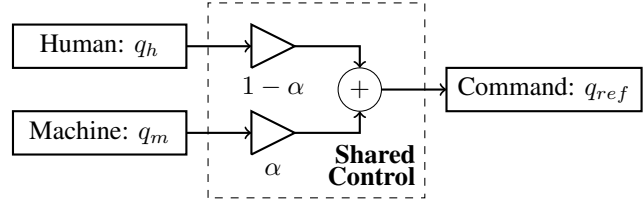
\begin{figure}[t!]
\centering
\begin{tikzpicture}[scale=0.47*\textwidth/29cm]


\draw [thick,->] (2,2)--(5,2);
\draw [thick] (5,1)--(5,3)--(7,2)--(5,1);

\draw [thick,->] (2,6)--(5,6);
\draw [thick] (5,5)--(5,7)--(7,6)--(5,5);

\draw [thick,->] (7,2)--(10,2)--(10,3);
\draw [thick,->] (7,6)--(10,6)--(10,5);

\draw (10,4) circle (1);
\node at (10,4) {$+$};

\draw [thick,->] (11,4)--(14,4);

\draw [thick] (-6,1) rectangle (2,3);
\draw [thick] (-6,5) rectangle (2,7);
\draw [thick] (14,3) rectangle (23,5);

\draw [dashed] (3,-1) rectangle (13,8);

\node at (-2,6) {Human: $q_h$};
\node at (-2,2) {Machine: $q_m$};
\node at (18.5,4) {Command: $q_{ref}$};

\node [below] at (6,5) {$1-\alpha$};
\node [below] at (6,1) {$\alpha$};

\node [left] at (13,1) {\bf Shared};
\node [left] at (13,0) {\bf Control};

\end{tikzpicture}
\caption{Arbitration: inputs from the human $q_h$ and the machine $q_m$ are blended to determine a single  shared-control input  $q_{ref}$ by using the level of autonomy $\alpha \in [0,1]$. }
\label{fig:sharedControl}
\end{figure}

Shared human-machine control requires an arbitration approach to select the relative amount of human and machine control. One  approach is for the human (or humans) to fully guide the machine as in teleoperation~\cite{Nudehi2005, Khademian2012}. Alternatively, the machine might facilitate human operation, e.g.,  through artificial potential fields~\cite{Khatib1986}  for lane keeping or hazard avoidance~\cite{Gerdes2001} and through virtual fixtures~\cite{Rosenberg1993} to guide task completion~\cite{OMalley2006}. 
In contrast to these human-centered or machine-centered approaches, blending of both the human and the machine input~\cite{Yu2003, Hansson2010} can leverage both the 
adaptability  of the human as well as the computing power and bandwidth of the machine controller. One approach to blend the human input $\bm{q_h}$ and the machine input $\bm{q_m} $ is through an adjustable level of autonomy (LOA) $\alpha \in [0,1]$~\cite{Yu2003},  e.g., to determine the reference input $\bm{q_{ref}}$ to the system (such as the desired  position of a robot)  
\begin{equation}
	\label{eqn:reference}
	\bm{q_{ref}} = \alpha \bm{q_m} + (1 - \alpha) \bm{q_h}, 
\end{equation}
as represented in Fig.~\ref{fig:sharedControl}. 
As opposed to a fixed level~\cite{You2012,Parasuraman1997}  or discrete levels of autonomy $\alpha$~\cite{Kortenkamp2000}, this work considers the more general shared-control case where 
the level of autonomy $\alpha$ is allowed to \emph{slide} continuously as in~\cite{Desai2005,Dias2008}, which has been 
used in applications such as active mobility devices~\cite{Yu2003,Wang2014}, control of multi-agent UAV systems~\cite{Franchi2012}, semi-autonomous operation of large manipulator arms~\cite{Hansson2010}, and active driver assistance systems for lane keeping~\cite{Saleh2013}.  

\begin{figure*}[t!]
	\centering
    \includegraphics[width=\textwidth]{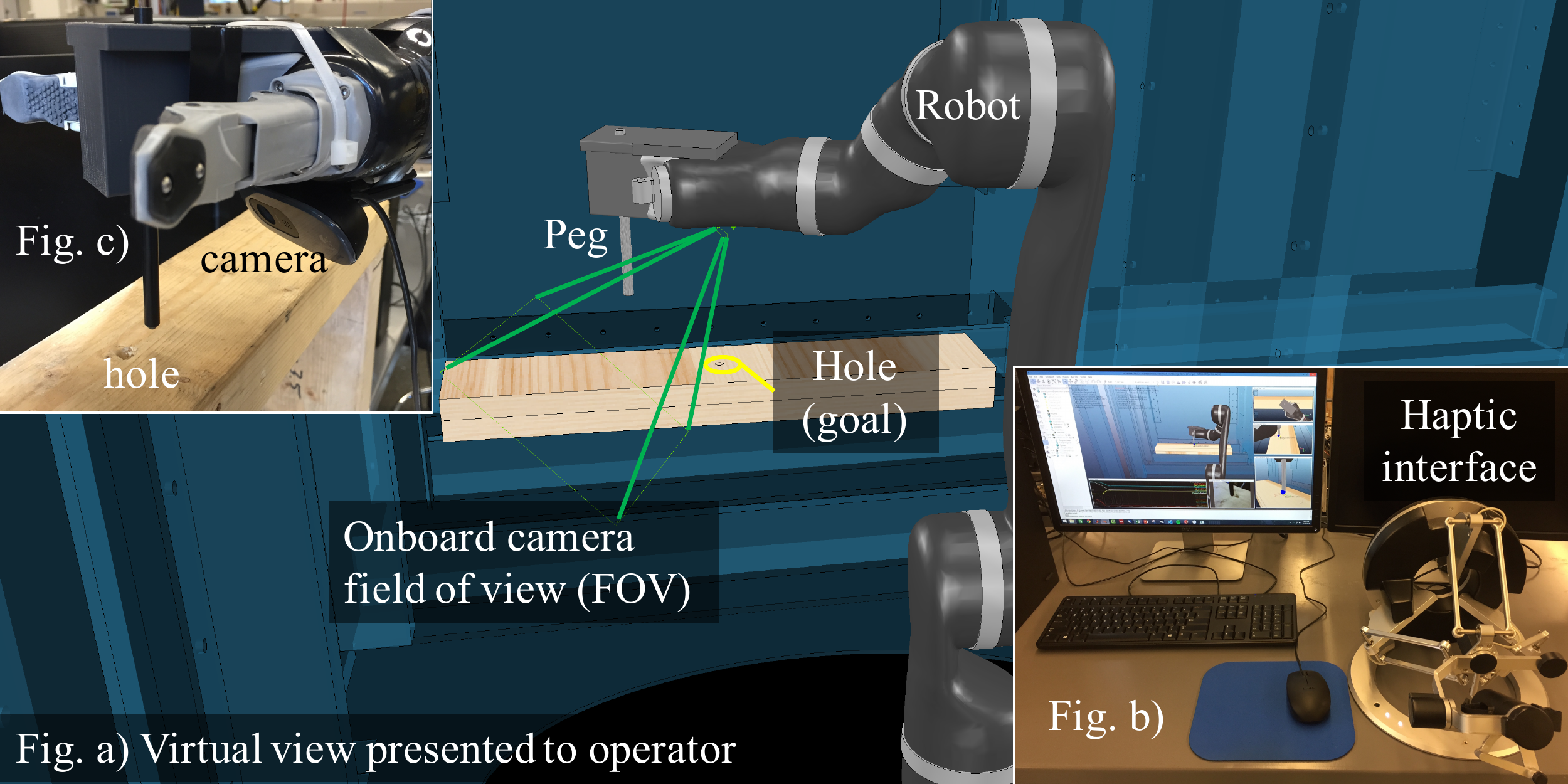}
    \caption{A user is a) presented a virtual image of the robot in the environment for a peg-in-hole task.  The estimated goal location ${\bm{\hat{q}_d}}$ is the center of the hole.  The user manipulates the peg using a haptic interface b) at the remote workstation.  Real camera feedback c) provides the operator with information of the actual goal location $\bm{q_d}$.}
    \label{fig:task}
\end{figure*}

\vspace{0.1in}
\par 
The main contribution of this  paper is a new approach to determine the level of autonomy for human-machine shared control based on the automation uncertainty, e.g., 
to reflect the level of machine confidence in the goal 
prediction, which can be used to determine when the automation can take over from a human~\cite{Dragan2013}. 
In aerospace manufacturing operations,  there can be  uncertainty in the location of the obstacles and the goal location for a manufacturing operation such as drilling.  If these uncertainties are substantial, then a human might need to take over the manufacturing operation. 
The  current work proposes a method for using apriori knowledge of automation uncertainty and the probability of failure to arbitrate the level of autonomy in shared control.  
For example as the probability $P(\mathcal{E})$ of the failure event  $\mathcal{E}$ increases,  the level of autonomy (LOA)  tends to zero, $\alpha \rightarrow 0$, and more control authority is relinquished to the human.  Conversely, the level of autonomy becomes larger, $\alpha \rightarrow 1$, when the probability of a failure event is low, giving more control authority to the machine.
Additionally, the proposed level of autonomy $\alpha$ is also used to scale the haptic feedback~\cite{Hannaford1989,Hogan1985}   assisting the user in task completion~\cite{Griffiths2004} to convey the level of automation confidence to the operator~\cite{Abbink2012}. 
Experimentation results, with a human-robot peg-in-a-hole testbed, show more than 5 times improvement in the error tolerance for task completion with the proposed shared control approach when compared to a purely autonomous method. 
In Section~\ref{sec:application} an application is described.  The control approach for the application is identified in Section~\ref{sec:control}.  In Section~\ref{sec:autonomySelection} machine confidence is generalized to principal modes of uncertainty, and a policy is derived for shifting authority based on environmental uncertainty.  Haptic virtual interaction forces are discussed in Section~\ref{sec:haptics}.  In Section~\ref{sec:results}, experimental results are shown comparing the novel shared control scheme with a purely autonomous process.


\begin{figure}[t!]
	\centering
    \includegraphics[width=0.48\textwidth]{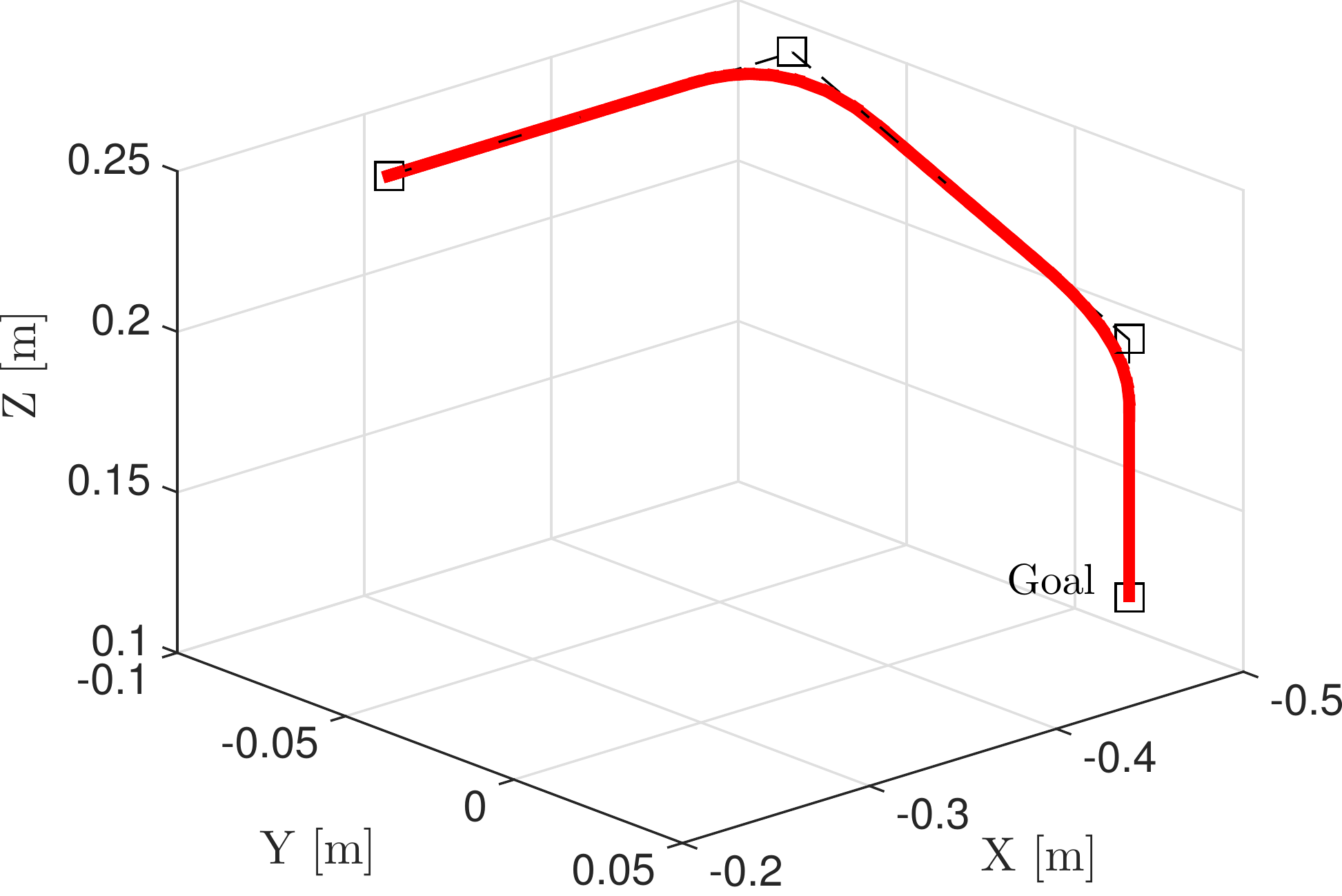}
    \caption{A smooth time-based automation trajectory $\bm{q_m}$ based on manual waypoints, Bezier curves, and acceleration and velocity limits.}
    \label{fig:machine}
\end{figure}

\begin{table}[t!]
\centering
\caption{Machine Trajectory}
\begin{tabular}{| c | c | c |} \hline
Parameter & Value & Units \\\hline\hline
Velocity at end-points & 0 & m/s \\\hline
Max acceleration & 2.0 & m/s$^2$ \\\hline
Max velocity & 0.2 & m/s \\\hline
\end{tabular}
\label{tbl:machine}
\end{table}

\begin{figure*}[t!]
\centering
\begin{tikzpicture}[scale=0.75*\textwidth/40cm]


\draw [thick,->,dashed] (21,3.5)--(19,3.5)--(12,10);

\draw [thick] ( 4, 9) rectangle (10,12);		
\draw [thick] ( 4, 3) rectangle (10, 6);		
\draw [thick,fill=white] (12, 6) rectangle (18, 9);		
\draw [thick] (21, 6) rectangle (27,10);		
\draw [thick] (21, 1) rectangle (27, 5);		
\draw [thick] (31, 6) rectangle (37, 9);		
\draw [thick] ( 4,13) rectangle (10,16);		
\draw [thick] (21,15) rectangle (27,19);		
\draw (0,  5) circle (.5);

\draw [thick,<->] (0,5.5) -- (0, 9.5) -- (4, 9.5);
\draw [thick,-]  ( -1,9.5) -- (  0,9.5);			
\draw [fill=black] (0,9.5) circle (.2);
\draw [thick,->] (0.5,  5) -- (  4,  5);			

\draw [thick,->] (  0, 2.5) -- (0, 4.5);			

\draw [thick,->] ( 10,11) -- ( 15, 11) -- ( 15, 9);	
\draw [thick,->] ( 12,11) -- ( 12, 15) -- ( 10, 15);
\draw [fill=black] (12, 11) circle (.2);

\draw [thick,->] (4,14.5)--(2,14.5)--(2,11.5)--(4,11.5);

\draw [thick,->] ( 10, 4.5) -- ( 15, 4.5) -- ( 15, 6);	
\draw [thick,->] ( 11, 4.5) -- ( 11,  14) -- ( 10,14);
\draw [fill=black] (11,4.5) circle (.2);

\draw [thick,->] ( 18,7.5) -- ( 21,7.5);				

\draw [thick,->] ( 27,8) -- ( 31,8);					

\draw [thick,->] ( 27,7)--(29,7)--(29,3.5)--(27,3.5);	

\draw [thick,->] (37,7.5)--(38,7.5)--(38,17.5)--(27,17.5);

\draw [thick,->] (24,12)--(24,10);
\draw [thick,->] ( 7, 1)--( 7,3);

\draw [thick,->] (28,8)--(28,16.5)--(27,16.5);
\draw [thick,->] (21,17)--(1,17)--(1,10.5)--(4,10.5);

\node at (7,10.5) {Human};
\node at (7, 5) {Machine}; \node at (7, 4) {Fig.~\ref{fig:machine}};
\node at (15,8) {Arbitration}; \node at (15,7) {Eqn.~\ref{eqn:reference}};
\node at (24,4) {Autonomy}; \node at (24,3) {Selection}; \node at (24,2) {Eqn.~\ref{eqn:filtPolicy}};
\node at (24,9) {Dynamics}; \node at (24,8) {Simulation}; \node at (24,7) {Fig.~\ref{fig:dynamics}};
\node at (34,7.5) {Robot};
\node at (7,15) {Haptics}; \node at (7,14) {Eqn.~\ref{eqn:haptics}};
\node at (24,18) {Visual}; \node at (24,17) {Feedback}; \node at (24,16) {Fig.~\ref{fig:task}};

\node [below] at (0, 2.5) {$\Delta_e$};
\node [left] at (-1,9.5) {$\bm{q_d}$};
\node [above] at (2,5) {$\bm{\hat{q}_d}$};

\node [left] at (0,6) {$+$};
\node [left] at (0,4) {$-$};

\node [right] at (18,5) {$\alpha$};

\node [above] at (14,11) {$\bm{q_h}$};
\node [below] at (14,4.5) {$\bm{q_m}$};

\node [above] at (19.5,7.5) {$\bm{q_{ref}}$};

\node [above] at (29.5,8) {$\bm{\theta_{ref}}$};
\node [right] at (29, 5) {$\hat{d}_e$};

\node [right] at (38,7.5) {$\bm{q}$};

\node [above] at (2.5,14.5) {$\bm{F_{\mathcal{H}}}$};

\node [below] at ( 7, 1) {$\hat{\mathcal{U}}$};
\node [above] at (24,12) {$\hat{\mathcal{U}}$};

\node [above] at (19,17) {$\mathcal{V}$};

\end{tikzpicture}
\caption{ 
In this paper, a human and machine share a common goal $\bm{q_d}$, but the machine is subject to an error $\Delta_e$ that influences the machine's estimate of the goal $\bm{\hat{q}_d}$. The human's input trajectory $\bm{q_h}$ is blended with the machine's input trajectory $\bm{q_m}$ through an arbitration scheme.  The resulting reference trajectory $\bm{q_{ref}}$ is passed through a dynamics simulation to generate the smoothened (dynamically viable)  robot joint angles $\bm{\theta_{ref}}$.  
The dynamics simulation also returns the  closest collision distance $\hat{d}_e$ from the robot tip $\bm{q_{ref}}$ to the nominal geometry of the environment $\hat{\mathcal{U}}$. 
When the robot is approaching the goal $\bm{q_d}$, this distance closely approximates the distance of the robot to the machine goal $\bm{\hat{q}_d}$.  The autonomy selection policy is used to adjust the level of autonomy $\alpha$ based on this distance $\hat{d}_e$. Visual feedback $\mathcal{V}$ of both the true robot position $\bm{q}$ (real camera feedback) and the simulated robot position (virtual reality) are provided to the human user.  In addition, haptic information about the machine trajectory $\bm{q}_m$ and the nominal environment $\hat{\mathcal{U}}$ (not shown) is fed back to the user in the form of interaction forces $\bm{F_\mathcal{H}}$.
}
\label{fig:comahi}
\end{figure*}

\section{Application and Assumptions}
\label{sec:application}

\subsection{Peg in a Hole Task}
A peg-in-hole task is studied in this work to illustrate the proposed uncertainty-based arbitration. A peg (representative of a drill bit)  fixed to the end of a serial manipulator as shown in Fig.~\ref{fig:task} is to be inserted into a hole at location $\bm{q_d}$ on a planar surface. A minimum of five degrees of freedom (DOF) endpoint control are needed for this task (3 DOF in translation and 2 DOF in normal orientation to a plane). The robot used is a Kinova MICO with 6-joints, sufficient for completing the peg-in-hole task.


\subsection{Machine Trajectory Input}
A time-dependent machine trajectory $\bm{q_m}(t)$ is constructed offline using waypoints and Bezier curves~\cite{Choi2014}, where the final point of the trajectory (at $t = t_f$) is the nominal goal location, i.e., $\bm{q_m}(t_f) = \bm{\hat{q}_d}$. The automation trajectory remains identical for the entire experiment.  Fig.~\ref{fig:machine} shows waypoints and the smoothed trajectory in workspace coordinates.  Parameters of the acceleration profile include starting velocity, velocity limits, and acceleration limits, summarized in Table~\ref{tbl:machine}.

\subsection{Uncertainty and Human Input}
\par Because the machine trajectory is planned for the nominal hole location $\bm{\hat{q}_d}$, a large error $\Delta_e$ in the actual hole location $\bm{q_d}$ could result in an automation failure.   While the actual location of the hole is unknown to the robot, the uncertainty in the hole location is assumed to be known as a Gaussian distribution $\bm{q_d} ~\sim \mathcal{N}(\bm{\hat{q}_d},\sigma_e^2)$, with the nominal hole location $\bm{\hat{q}_d}$ as in Fig.~\ref{fig:task}a, and variance of the distribution is $\sigma_e^2$.    The error of the actual hole location versus the nominal hole location is $\Delta_e = \bm{q_d} - \bm{\hat{q}_d}$.
To manage this error in the hole location, human input is included in the control loop.  The human input $\bm{q_h}$ is received via a Force Dimension haptic interface, shown in Fig.~\ref{fig:task} b).  The operator input $\bm{q_h}$ is combined with the machine input $\bm{q_m}(t)$ according to~\eqref{eqn:reference}.

\subsection{Human Interface}
The human operator is presented with virtual interaction forces via the Force Dimension haptic display.  These forces include a virtual stiffness centered around the machine trajectory as well as the obstacles in the environment.  The haptic display does not include measured forces from the robot.  The platform used for providing haptic rendering is the CHAI3D library, operating at 2 kHz.

\par Real camera feedback of the actual environment at the end of the arm, shown in Fig.~\ref{fig:task} c), is provided to the user real-time.  The hole is in the field of view (FOV) when the robot is near the goal $\bm{q_d}$.  This view allows for the human to close the loop around the true goal.  Virtual views of the robot in the environment are also presented to the user, similar to Fig.~\ref{fig:task} a).

\par The operator interface and high-level control is handled by Virtual Robotic Experimentation Platform (V-REP) software at a workstation isolated from the operating environment of the robot.  The workstation and software are shown in Fig.~\ref{fig:task} b).  This software runs at $\sim$20 Hz, and provides a virtual simulation of the robot in the environment.  V-REP also provides a positive estimated scalar distance $\hat{d}_e$ of the robot end-effector to the nearest obstacle or environmental surface, available via nominal CAD (computer aided design) data $\hat{\mathcal{U}}$. This CAD data contains the feature location of the nominal hole $\bm{\hat{q}_d}$.  When the robot is near the goal, the distance $\hat{d}_e$ approximates the distance of the peg to the nominal hole $\bm{\hat{q}_d}$.

\section{Control Approach}
\label{sec:control}

In human machine interaction, shared control is formulated as automation sharing control with a human.  Fig.~\ref{fig:comahi} shows a block diagram for the control approach for this paper.  In this experiment, it is assumed that the machine and human share a common goal $\bm{q_d}$.  
Machine error $\Delta_e$ is introduced into the system as an automation error due to environment uncertainty.  To keep hardware and sensor costs practical, knowledge of the behavior of this error is used to determine when to incorporate human intervention, e.g. when an error $\Delta_e$ in the environment is high, and as such the automation cannot complete a task, human adaptability can be incorporated in the control.

\par
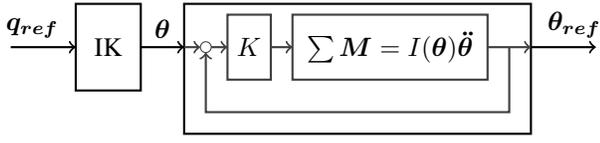
\begin{figure}[t!]
\centering
\begin{tikzpicture}[scale=0.47*\textwidth/29cm]


\draw [thick,->] (0,4)--(3,4);
\node [above] at (1,4) {$\bm{q_{ref}}$};

\draw [thick] (3,2) rectangle (6,6);
\node at (4.5,4) {IK};

\draw [thick,->] (6,4)--(8,4);
\node [above] at (7,4) {$\bm{\theta}$};

\draw [thick] (8,0) rectangle (24,6);

\draw [thick,color=darkgray,->] (8,4)--(8.75,4);
\draw [color=darkgray] (9,4) circle (.25);
\draw [thick,color=darkgray,->] (9.25,4)--(10,4);
\draw [thick,color=darkgray] (10,2.5) rectangle (12,5.5);
\node at (11,4) {$K$};
\draw [thick,color=darkgray,->] (12,4)--(13,4);
\draw [thick,color=darkgray] (13,2.5) rectangle (22,5.5);
\node at (17.5,4) {$\sum \bm{M} = I(\bm{\theta})\bm{\ddot{\theta}}$};
\draw [thick,color=darkgray,->] (22,4)--(24,4);
\draw [thick,color=darkgray,->] (23,4)--(23,1)--(9,1)--(9,3.75);

\draw [thick,->] (24,4)--(27,4);
\node [above] at (26,4) {$\bm{\theta_{ref}}$};

\end{tikzpicture}
\caption{A dynamics simulation step shapes the reference trajectory of manipulator joint angles via Inverse Kinematics (IK) based on a simple torque control loop $K$ and manipulator dynamic properties $\sum \bm{M} = I(\bm{\theta})\bm{\ddot{\theta}}$. This dynamic simulation filters commanded joint angles $\bm{\theta_{ref}}$ sent to the physical robot.}
\label{fig:dynamics}
\end{figure}

 
The blended input trajectory $\bm{q_{ref}}$ is passed through the Inverse Kinematics (IK) of the robot model, using a damped least squares method~\cite{Wampler1986},  to generate a set of desired joint angles $\bm{\theta}$. These desired joint angles $\bm{\theta}$ are then 
passed through a nonlinear  dynamics simulation of the torque-controlled robot, as shown in Fig.~\ref{fig:dynamics}, to generate the joint commands $\bm{\theta_{ref}}$. 
This simulation tends to filter high frequency commands, and provides a means to generate a set of feasible joint commands $\bm{\theta_{ref}}$.  
Closed-loop motor control to achieve the reference trajectory $\bm{\theta_{ref}}$ is handled by the software provided with the robot.  When automation uncertainties are present in the system, the human input $\bm{q_h}$ provides closed-loop control based on visual feedback $\mathcal{V}$ to correct for errors $\Delta_e$ in the estimated goal location $\hat{\bm{q_d}}$.


\subsection{Shared Control}
\begin{figure}[t!]
	\centering
    \begin{tikzpicture}[scale=0.47*\textwidth/26cm,
declare function={ alpha(\x)    = 1 - exp( - pow(\x-20,2)/(2*pow(4,2)) );
				   mTraj(\x)  = 10 - 0.2*\x + 0.03*\x*\x - 0.001*\x*\x*\x;
			       hTraj(\x)    = 9 - 0.4*\x + 0.025*\x*\x - 0.0005*\x*\x*\x;
},
]


\draw[<->,gray] (0,11) -- (0,6) -- (22,6);
\node [right,gray] at (22,6) {$t$};
\node [above,gray] at (0,11) {$q$};

\draw[gray] (15.15,{mTraj(15.15)}) -- (15.15,{alpha(15.15)* mTraj(15.15) + (1 - alpha(15.15))* hTraj(15.15)});
\draw[gray] (15.15,{hTraj(15.15)}) -- (15.15,{alpha(15.15)* mTraj(15.15) + (1 - alpha(15.15))* hTraj(15.15)});
\node [right] at (15.15,9.5) {$(\alpha) q_m$};
\node [left] at (15.15,8) {$(1 - \alpha) q_h$};
\draw[fill,darkgray] (15.15,{mTraj(15.15)}) circle [radius=0.07];
\draw[fill,darkgray] (15.15,{hTraj(15.15)}) circle [radius=0.07];
\draw[fill,darkgray] (15.15,{alpha(15.15)* mTraj(15.15) + (1 - alpha(15.15))* hTraj(15.15)}) circle [radius=0.07];

\draw[domain= 0:15,smooth,blue,->,ultra thick] plot ({\x},{hTraj(\x)});
\draw[domain=15.3:20,smooth,blue,-,ultra thick] plot ({\x},{hTraj(\x)});
\node[blue,left] at (5,7) {$q_h$};

\draw[domain= 0:15,smooth,red,->,ultra thick] plot ({\x},{mTraj(\x)});
\draw[domain=15.3:20,smooth,red,-,ultra thick] plot ({\x},{mTraj(\x)});
\node[red,above] at (20,10) {$q_m$};

\draw[domain= 0:15,smooth,teal,->,ultra thick] plot ({\x},{alpha(\x)* mTraj(\x) + (1 - alpha(\x))* hTraj(\x)});
\draw[domain=15.3:20,smooth,teal,ultra thick] plot ({\x},{alpha(\x)* mTraj(\x) + (1 - alpha(\x))* hTraj(\x)});
\node[teal,above] at (3,9.5) {$q_{ref}$};

\draw[fill] (20,{hTraj(20)}) circle [radius=0.25];
\node [right] at (20,{hTraj(20)+.5}) {Goal: $q_d$};

\draw[<->,gray] (0,4) -- (0,0) -- (22,0);
\node [right,gray] at (22,0) {$t$};
\node [above,gray] at (0,4) {$\alpha$};
\node [left,gray] at (0,0) {$0$};
\node [left,gray] at (0,3) {$1$};

\draw[domain= 0:15,smooth,->,thick] plot({\x},{3*alpha(\x)});
\draw[domain=15.2:20,smooth,thick] plot({\x},{3*alpha(\x)});

\node[below] at (4,0) {Autonomous};
\node[below] at (12,0) {Semi-}; \node[below] at (12,-1) {autonomous};
\node[below] at (19,0) {Tele-}; \node[below] at (19,-1) {operation};

\end{tikzpicture}
    \caption{A 1D conceptual diagram shows sliding mode shared control varies the level of autonomy $\alpha$ continuously over time based on an autonomy selection criteria.  The approach in this work uses knowledge of machine uncertainties to vary the arbitration regime based on distance from the goal.}
    \label{fig:trajectory}
\end{figure}
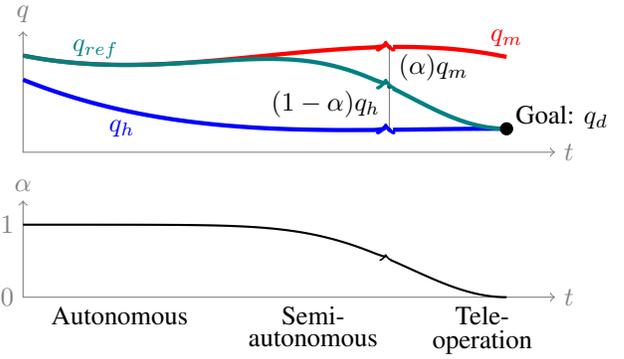

The human input $\bm{q_h}$ is applied through a haptic input device.  The machine input $\bm{q_m}$ is queried from the path planning module in Fig.~\ref{fig:machine}.  An autonomy selection policy defines the level of autonomy $\alpha$.  Fig.~\ref{fig:trajectory} shows how the reference trajectory changes as the level of autonomy (LOA) slides from a fully autonomous mode to a tele-operation mode.

\subsection{Autonomy Selection Policy}
\label{sec:autonomySelection}
Previous works~\cite{Dragan2013} have used the confidence interval in a machine prediction task as the level of autonomy $\alpha$, e.g.
\begin{equation}
	\label{eqn:dragan}
	\alpha = \max \left(0, 1 - \frac{d}{D}\right),
\end{equation}
where $d$ is the distance to a goal and $D$ is a threshold defining the point of $\alpha = 0$, i.e. no confidence in the machine.  This concept is extended in this work to define a general autonomy selection based on uncertainty in a task. 
In particular, if $P(\mathcal{E})$ is the probability of a machine failure mode $\mathcal{E}$ occurring, then the level of autonomy $\alpha$ is chosen as  
\begin{equation}
	\label{eqn:probAlpha}
	\alpha ~= 1-P(\mathcal{E}),
\end{equation}
where $P(\mathcal{E})$ is the probability of an environmental machine failure occurring. As the probability of the failure event increases, $\alpha \rightarrow 0$, i.e. more control authority is relinquished to the human.  Conversely, the level of autonomy $\alpha \rightarrow 1$ when the probability of failure event is low, giving more control authority to the machine.  

\subsubsection{Failure Estimation for Arbitrating Level of Autonomy}
\label{sec:arbitrationDerivation}
%


Consider the 1-dimensional case of the robot tip  $q_t$ approaching a goal $q_d$, with initial $q_t > q_d$.  The nominal goal is located at $\hat{q}_d$.  From section~\ref{sec:application}, the probability density function (pdf) of the goal location $q_d$ is normal, i.e., $q_d \sim \mathcal{N}(\hat{q}_d,\sigma_e^2)$, and given by 
\begin{equation}
	\label{eqn:envrPDF}
	f(q_d) = \frac{1}{\sigma_e \sqrt{2\pi}} \exp \left\{ -\frac{(q_d - \hat{q}_d)^2}{2\sigma_e^2} \right\}.
\end{equation}
The complementary cumulative distribution function $\bar{F}(q)$ is
\begin{equation}
	\bar{F}(q) = \int_q^{\infty} f(s) ds = \left. -\frac{1}{2} \erf \left\{ \frac{\hat{q}_d - s}{\sigma_e \sqrt{2}} \right\} \right|_q^{\infty},
\end{equation}
which reduces to
\begin{equation}
	\bar{F}(q) = \frac{1}{2} \left[1 + \erf \left\{ \frac{\hat{q}_d - q}{\sigma_e \sqrt{2}} \right\} \right].
\end{equation}
Since the robot tip approaches with  $\hat{q}_d >  q$,  the probability $P(\mathcal{E})$ of encountering the hole at $q_t$ is the complementary cumulative distribution $\bar{F}(q_t)$
\begin{equation}
	\label{eqn:cdfTip}
	\begin{array}{rcl}
	P(\mathcal{E}) ~= \bar{F}(q_t) & = &  \frac{1}{2} \left[1 + \erf \left\{ \frac{\hat{q}_d - q_t}{\sigma_e \sqrt{2}} \right\}\right] \\[0.5em]
	& = & \frac{1}{2} \left[ 1 - \erf \left\{ \frac{\hat{d}_e}{\sigma_e \sqrt{2}} \right\}\right]
	\end{array}
\end{equation}
where $\hat{d}_e = q_t - \hat{q}_d$ is the measured distance of the robot tip $q_t$ from the nominal goal location $\hat{q}_d$. 
Note that if the goal is located directly on a geometric surface (e.g., a wall or obstacle), then the probability of encountering the goal is the probability of encountering an obstacle.
\par 
From \eqref{eqn:probAlpha} and \eqref{eqn:cdfTip}, the level of autonomy $\alpha$ can be rewritten as 
\begin{equation}
	\label{eqn:policy}
	\alpha = 1 - P(\mathcal{E}) = \frac{1}{2}\left[1 + \erf \left\{ \frac{\hat{d}_e}{\sigma_e \sqrt{2}} \right\}\right].
\end{equation}
In this work, the robot approaches the goal from above (orthogonal to the surface in which the hole is located), so the probability of encountering the surface is $P(\mathcal{E})$, where the error in hole location is in the $z$ direction, i.e., $\Delta_{e_z}$.  The robot will collide with the surface if the hole centerline is off by some error $\Delta_{e_x}$, so experimentally $\Delta_{e_z}$ and $\Delta_{e_x}$ must be varied to evaluate~\eqref{eqn:policy}.
\par 
This representation allows the level of autonomy $\alpha$ to slide toward human intervention when the likelihood of encountering the goal (and potential failure due to uncertainty in the goal location) is high.  This is different from typical cases where 
the level of autonomy $\alpha$ shifts toward machine control when the distance to the goal $\hat{d}_e$ is small because the robot becomes confident in its prediction of the goal~\cite{Dragan2013}.

\par An advantage of the approach in~\eqref{eqn:probAlpha} is that
in cases with multiple independent failure modes $\mathcal{E}_i$ due to uncertainty, the selection of the level of autonomy $\alpha$ can be generalized as 
\begin{equation}
	\label{eqn:prob}
	\alpha = 1 -  \prod_i P({\mathcal{E}}_i).
\end{equation}
Thus, the proposed uncertainty-based approach can be used to 
manage different uncertainty modes in shared human-machine control such as: (i) \emph{environmental uncertainty} that arises when there is a change in an environment from the apriori model such as tolerance variation in an assembly from its geometric design; (ii) \emph{method uncertainty} that occurs when automation fails to find a solution to a task such as a path planning module failing to find a trajectory; and (iii) \emph{measurement uncertainty} arises when a sensor has substantial errors due to external noise. 

\begin{figure}[t!]
	\centering
    \includegraphics[width=0.48\textwidth]{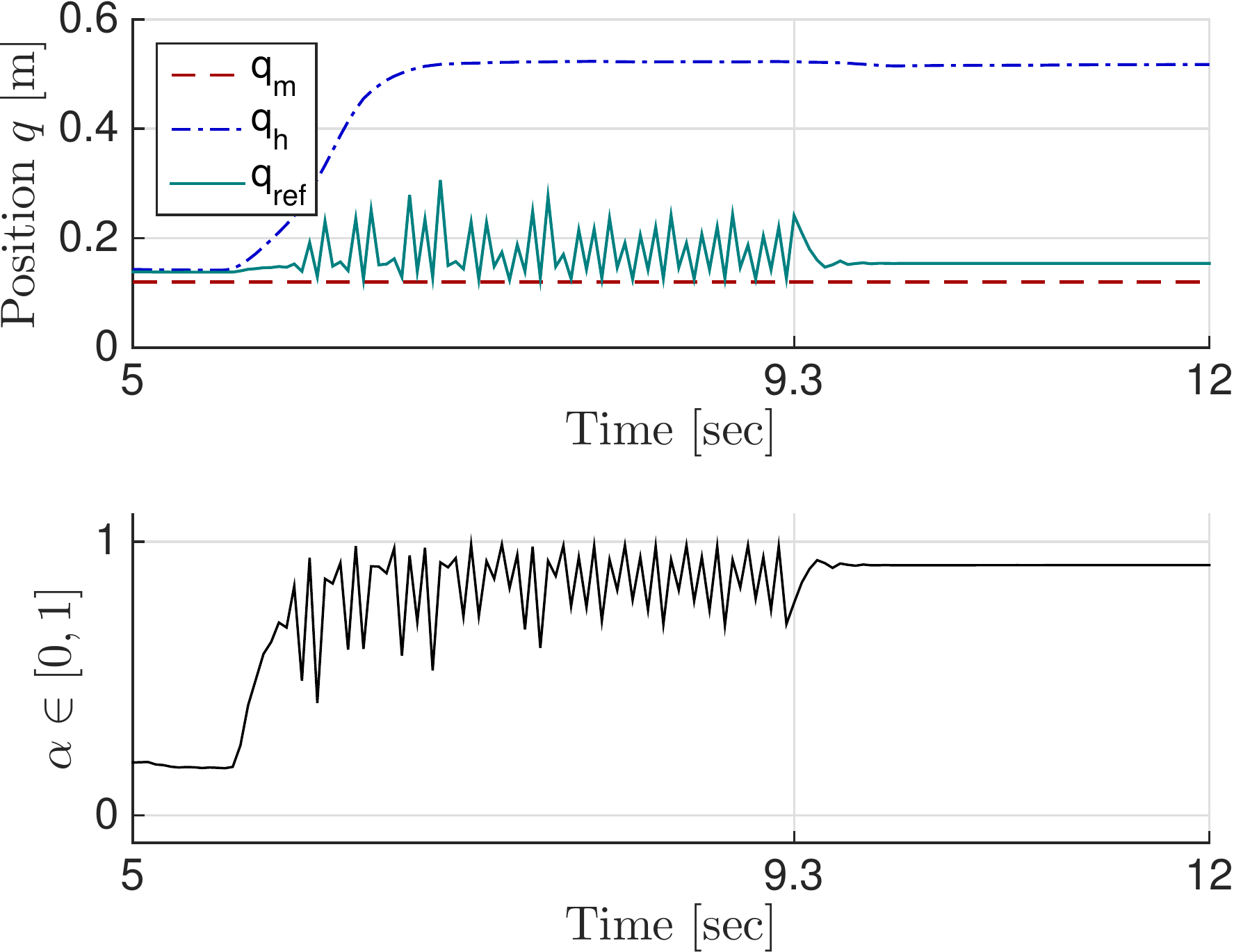}
    \caption{Without the filter in~\eqref{eqn:filtPolicy}, a chattering effect on the level of autonomy $\alpha$ is noticeable in experimentation when the difference between human input and machine input $|\bm{q_h} - \bm{q_m}|$ is relatively large.  Enabling a filter in~\eqref{eqn:filtPolicy} on the level of autonomy $\alpha$ at 9.3 seconds removes the chattering phenomenon.}
    \label{fig:chattering}
\end{figure}

\subsubsection{Chattering effect}
Rapid switching  between the human control mode and the machine control mode is possible close to the location of a potential failure. This is 
similar to chattering found in sliding mode control (SMC) architectures~\cite{Young1999}.  Fig.~\ref{fig:chattering} shows experimental data with chattering for level of autonomy $\alpha$ defined by~\eqref{eqn:policy}.  The chattering mode is excited in experimentation when: (i)~there is a  large difference between human input and machine input $|\bm{q_h} - \bm{q_m}|$ and (ii)~the distance to the nominal goal location $\hat{d}_e$ is close to  variance $\sigma_e$ of the uncertainty in the goal location.

\par The chattering effect is reduced by passing changes in the level of autonomy $\alpha$ through a lowpass  (first order) filter, i.e., 
\begin{equation}
	\label{eqn:filtPolicy}
	\xi \dot{\alpha} + \alpha = \frac{1}{2}\left[1 + \erf \left\{ \frac{\hat{d}_e}{\sigma_e \sqrt{2}} \right\}\right],
\end{equation}
where $\xi$ is the filter time constant.   
The filter time constant  $\xi$ is chosen to be $0.08$s, which is just faster than a the  human  neuromuscular reflex time constant of 0.1 s~\cite{Saleh2013}. Fig.~\ref{fig:chattering} presents experimental results that show that enabling the filter  (at $t = 9.3$s) mitigates the chattering phenomenon.  

\subsection{Haptics Generator}
\label{sec:haptics}
%
%
The haptic interaction force $\bm{F_{\mathcal{H}}}$ applied to the user is the linear combination of virtual fixtures $\bm{F_f}$ for guiding the operator along the machine trajectory~\cite{Rosenberg1993} and potential fields $\bm{F_v}$ for keeping the operator input away from collisions~\cite{Khatib1986}
\begin{equation}
	\bm{F_{\mathcal{H}}} =  \bm{F_f}(\alpha,\bm{q_h},\bm{\dot{q}_h}) + \bm{F_v}(\alpha,\bm{q_h}).
	\label{eqn:haptics}
\end{equation}

\subsubsection{Virtual Fixtures}
\par A virtual fixture comprised of variable linear stiffness $k$ and viscous damping $b$ is incorporated into the system.  The virtual fixture force $\bm{F_f}$ applied to the user is
\begin{equation}
	\bm{F_f} = -k(\bm{q_h} - \bm{q_m}) - b \bm{\dot{q}_h},
\end{equation}
where $\bm{q_h}$ is the input from the human, and $\bm{q_m}$ is the input from the machine.  Griffiths and Gillespie's premise of haptically conveying machine confidence~\cite{Griffiths2005} is accomplished by varying stiffness $k$ as a function of level of autonomy $\alpha$
\begin{equation}
	\label{eqn:stiffness}
	k = \alpha(k_{max}-k_{min})+k_{min},
\end{equation}
where $k_{min}$ and $k_{max}$ are the minimum and maximum stiffness values imposed on the system, respectively. Using the arbitration factor $\alpha$ to scale stiffness resolves trade-off issues associated with parameter selection for virtual fixtures.  In cases of high certainty ($\alpha \rightarrow 1$), the virtual fixture is strongly imposed on the user.  Conversely, in cases of low certainty ($\alpha \rightarrow 0$), the virtual fixture is weakly imposed.

\subsubsection{Selection of Haptics Parameters}
%
%
The haptic feedback gains $k_{min}$, $k_{max}$, and $b$ are selected experimentally as in~\cite{Marayong2004}.   A user familiar with operating the haptic interface moves a cursor to a visual goal constrained by different haptic gains.  These gains are held constant for the duration of each test, and parameters are evaluated for a no goal error case and a large goal error case, similar to Marayong and Okamura's method.

\par
\begin{figure}[t!]
	\centering
    \includegraphics[width=0.48\textwidth]{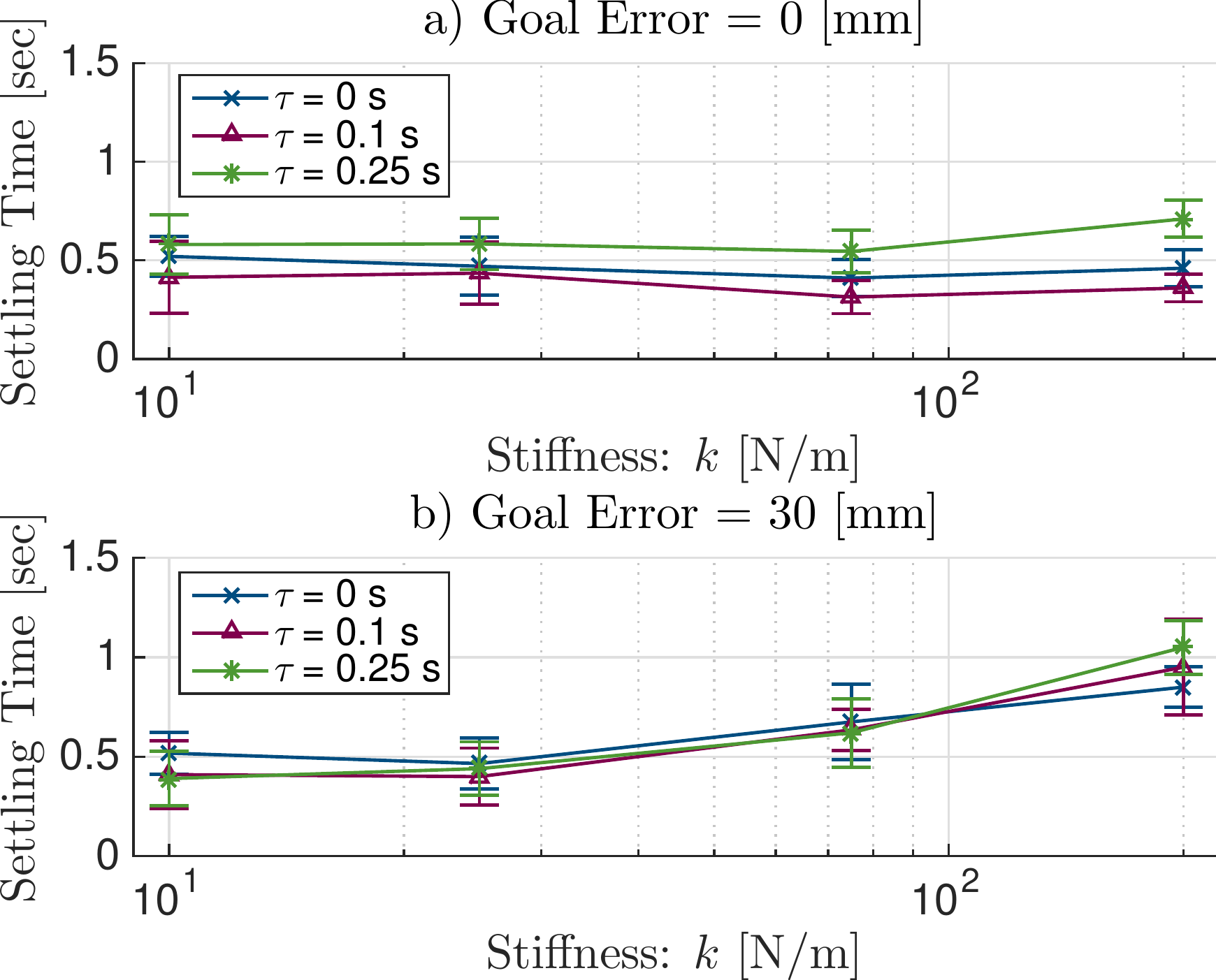}
    \caption{Settling time response characteristics are shown for parameter variations in haptic time constant $\tau = b/k$ and stiffness $k$ for a) cases with no goal error, and b) cases with a large goal error $\Delta$ of 30 mm.  Error bars show standard deviation for a minimum of 10 runs.}
    \label{fig:parameters}
\end{figure}

 The settling time for different selections of stiffness parameters $k_{max}$, $k_{min}$, and damping $b$ is shown in Fig.~\ref{fig:parameters}.
The parameter $k_{min}=10$N/m was selected to  minimize the settling time $T_s$ for a large goal error $\Delta = 30$mm, and 
the parameters $k_{max}=75$~N/m and $\tau = b/k_{max} =0.1$s were selected to  minimize the  settling time $T_s$ with no goal error $\Delta = 0$~mm as seen from 
Fig.~\ref{fig:parameters}. This results in a constant damping $b= \tau k_{max} = 7.5$~N-m/s.

\subsubsection{Potential Fields}
\par Potential fields $\bm{F_v}$ are also implemented as obstacles in the haptic workspace with a variable stiffness $k_v$ normal to the obstacle mesh plane $\bm{q_{obs}}$, such that
\begin{equation}
	\bm{F_v} = -k_v(\bm{q_h} - \bm{q_{obs}}),
\end{equation}
and $k_v$ is generated by
\begin{equation}
	\label{eqn:stiffness2}
	k = \alpha(k_{v_{max}}-k_{v_{min}})+k_{v_{min}},
\end{equation}
where $\alpha$ is the level of autonomy, and $k_{v_{max}}$ and $k_{v_{min}}$ are maximum and minimum virtual fixture stiffness, respectively.  Unless contact is made with these potential field regions, this force is not felt by the user.  The goal of the potential field is to show the user where obstacles might be, and although the user may need to overcome the field, it should less compliant than the virtual fixture.  The maximum stiffness for the haptic device before it exhibits instabilities was experimentally found to be  $k_{v_{max}} = 1000$~N/m.  A minimum stiffness $k_{v_{min}}$ of $200$~N/m was selected to provide a guiding field that the user can easily perceive but also overcome. 
Table~\ref{tbl:haptics} shows the virtual fixture parameters used in the experiments.

\begin{table}[t!]
\centering
\caption{Selected Haptics Parameters}
\begin{tabular}{| c | c | c | c |} \hline
Parameter & & Value & Units \\\hline\hline
Virtual Fixture Max Stiffness & $k_{max}$ & 75.0 & N/m \\\hline
Virtual Fixture Min Stiffness & $k_{min}$ & 10.0 & N/m \\\hline
Virtual Fixture Damping & $b$ & 7.5 & N-s/m \\\hline
Potential Field Max Stiffness & $k_{v_{max}}$ & 1000 & N/m \\\hline
Potential Field Min Stiffness & $k_{v_{min}}$ & 200 & N/m \\\hline
\end{tabular}
\label{tbl:haptics}
\end{table}

\section{Experimental Results}
\label{sec:results}
The proposed shared approach is compared to the case without shared control by experimentally evaluating the task completion time and the task success rate. 
In the first case, the machine autonomously attempts to complete the task with no operator intervention, i.e., without the shared control.  In the second case, the proposed shared control approach is used, and the operator can intervene to complete the task.  
In both instances, an error $\Delta_{e_z}$ between the actual $\bm{q_d}$ and nominal hole $\hat{\bm{q_d}}$ location height is generated randomly, according to the normal distribution assumed in Section~\ref{sec:application}, with a standard deviation of $\sigma_e = 10$~mm. 
To exercise the surface collision event considered in Section~\ref{sec:arbitrationDerivation}, 7 discrete error values in the hole centerline $\Delta_{e_x}$ are spread from $0$ to $3\sigma_e$ mm. These values are randomized to minimize human anticipation, with $N$ = 10 runs to capture completion time variance and success rate.  A user familiar with operating the system performed 70 runs of the semi-autonomous case, and data was collected for 70 runs of the purely autonomous case.

\begin{figure}[t!]
	\centering
    \includegraphics[width=0.48\textwidth]{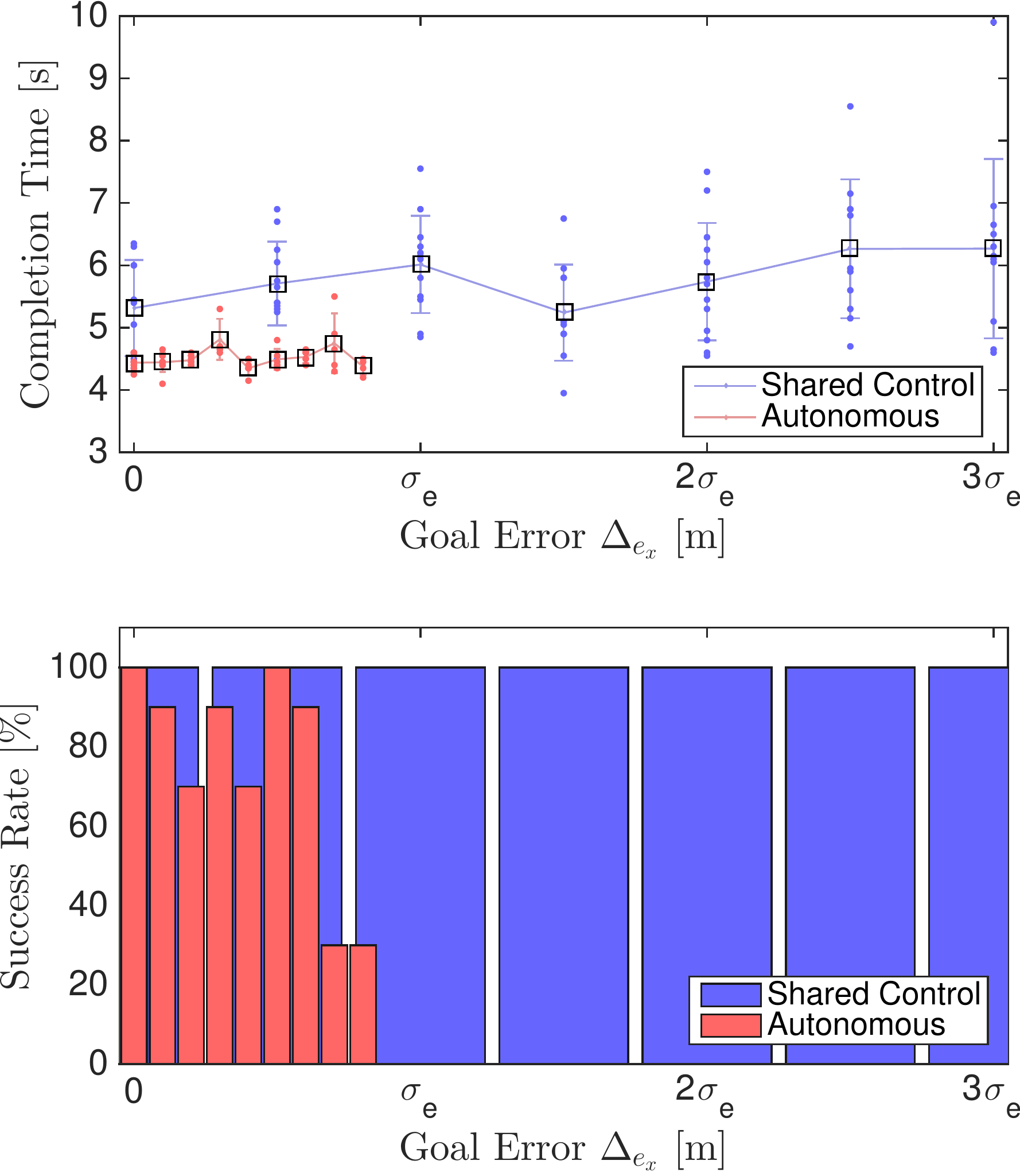}
    \caption{Experimental Results comparing task time and success rate to tolerance error for both pure automation (red) and the uncertainty-based shared control scheme (blue).}
    \label{fig:ph_results}
\end{figure}

\begin{figure}[t!]
	\centering
    \includegraphics[width=0.48\textwidth]{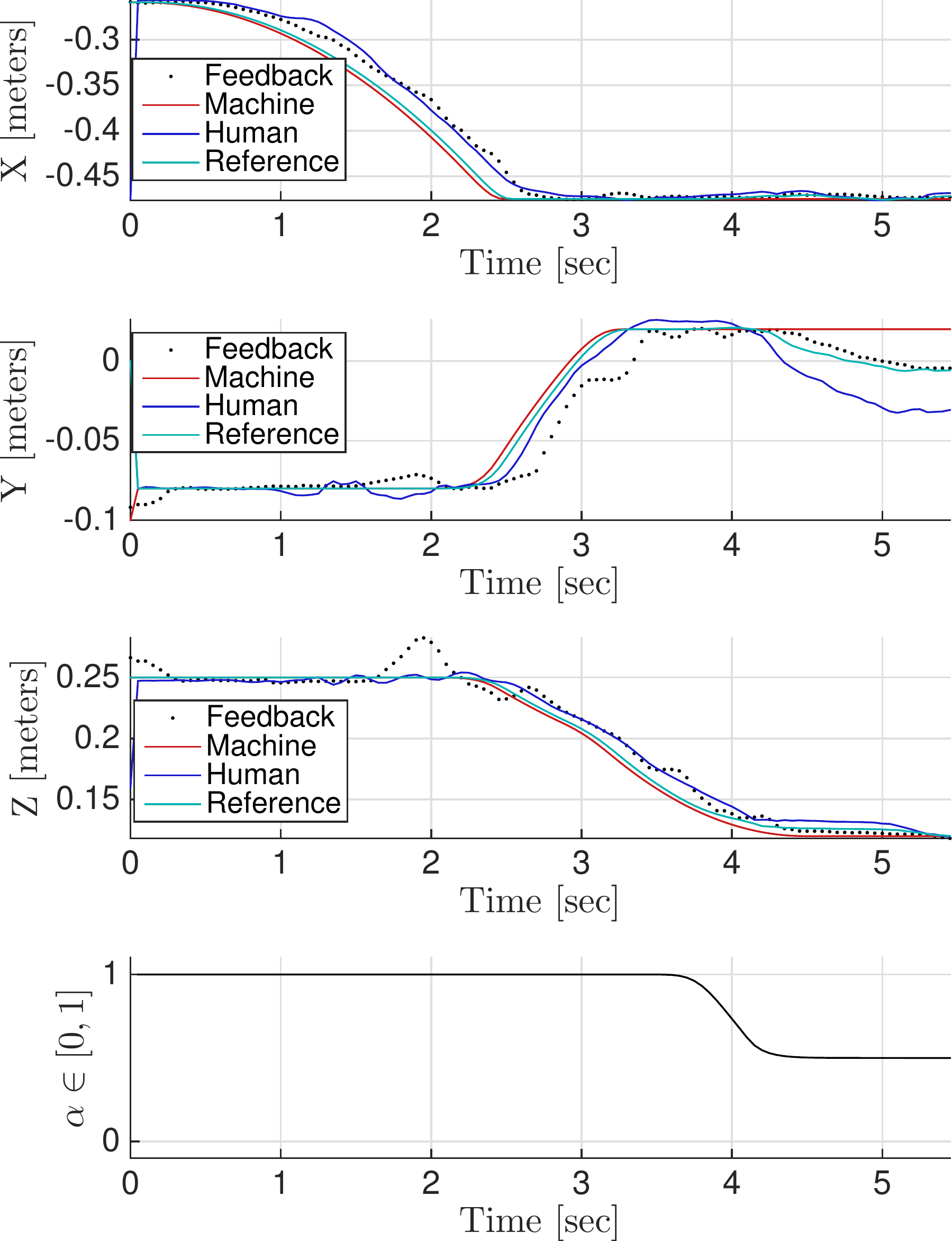}
    \caption{Experimental results showing time trajectories of the peg with  the shared control approach.  Notice the actual goal location is offset from the nominal goal
    location.}
    \label{fig:timehistdata}
\end{figure}

Completion times, with and without the shared control, are shown in Fig.~\ref{fig:ph_results}  with 10 runs for each data point.  Error bars show $\pm$ 1 standard deviation for indicated data points.  A typical time history of the peg for a large goal error ($\Delta_{e_x} = 20$ mm) is shown in Fig.~\ref{fig:timehistdata} for the case with shared control.  Initially, the human input (blue) closely follows the machine input (red) due to haptic interaction forces asserted on the operator.  Notice the slide in level of autonomy toward human authority ($\alpha \rightarrow 0$) as the tool tip reference $q_{ref}$ (cyan) begins to near the goal around $t > 3.5$ s.  During this phase, haptic virtual fixture forces are strongly imposed on the operator to keep the haptic tooltip following the machine trajectory.  The hole is not in the feedback camera field of view. At about $t = 4.2$ s, the operator realizes that the machine trajectory is incorrect 
via the camera feed, and uses this visual feedback to adjust the trajectory accordingly.  Because the robot is near an obstacle, the virtual fixture forces are only weakly imposed on the operator.  The potential field around the environment is also felt by the user, and the operator corrects for the incorrect machine trajectory by overcoming this field.  The peg is plugged into the hole successfully at $t = 5.4$ s.

\begin{figure}[t!]
	\centering
    \includegraphics[width=0.48\textwidth]{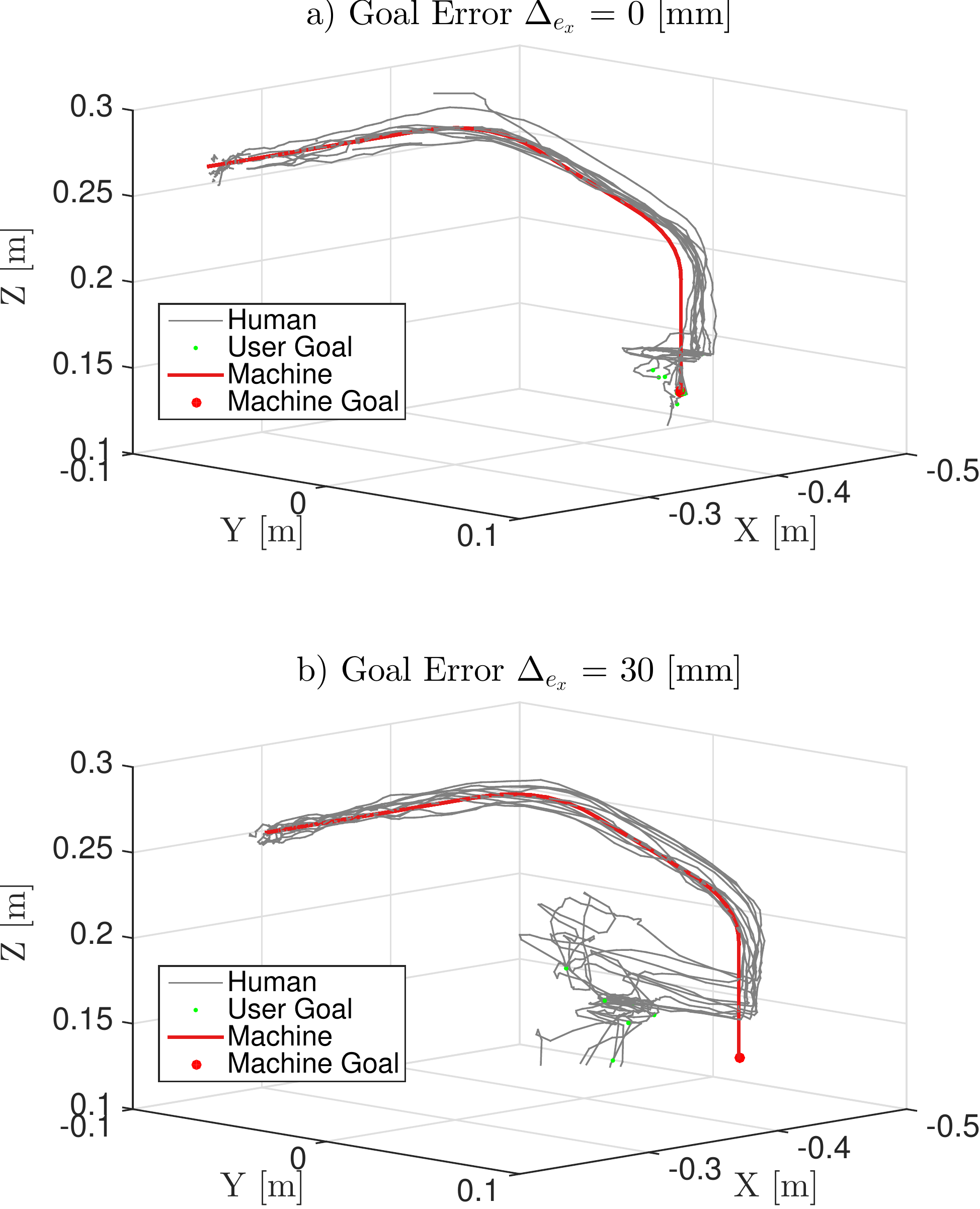}
    \caption{Experimental Results showing human $q_h$ and machine $q_m$ trajectories for a) no goal error $\Delta_e = 0$ mm and b) a large goal error $\Delta_e = 30$ mm.  Haptic interaction forces allow the operator input to closely track the machine input when a probability of collision is low, but sliding the level of autonomy gives the operator control authority when adaptation is needed.}
    \label{fig:3dtraj}
\end{figure}

Fig.~\ref{fig:3dtraj} compares shared control trajectories for no goal error and a large goal error.  With no goal error, the human is quickly able to complete the task since the true goal aligns with the estimated goal of the machine.  In the case of a large goal error however ($\Delta_{e_x} = 30$ mm), the operator realizes the machine trajectory is incorrect and compensates via visual feedback.

Mean completion time for the the purely autonomous system over all goal errors was $4.52$~s, with a standard deviation of $0.18$~s.  Mean completion time for the shared control method over all goal errors was $5.79$~s with a standard deviation of $0.93$~s.  A successful task is registered when the robot inserts the peg into the hole.  The autonomous system began to register more failed tasks than successes for goal errors $\Delta_{e_x} > 6$ m, while the shared control method had 100 \% successes over the entire experiment space as summarized in Table~\ref{tbl:results}.

\begin{table}[t!]
\centering
\caption{Experimental Results Summary}
\begin{tabular}{| c | c | c | c |} \hline
 & Autonomous & Shared Control & Units \\\hline\hline
Error $\Delta_{e_x}$ tolerance & 6 & 30 & mm \\\hline 
Mean Completion Time & 4.52 & 5.79 & s \\\hline
Completion Time StDev & 0.18 & 0.93 & s \\\hline
\end{tabular}
\label{tbl:results}
\end{table}

\section{Conclusion}
\label{sec:conclusion}
The probabilistic approach presented in this paper offers a method for incorporating human intervention in the case of automation uncertainty to increase task success with minimal impact to completion time.  This is accomplished by using apriori knowledge of uncertainties to arbitrate human-machine shared control.  Furthermore, haptic interaction virtual forces are scaled by the arbitration factor to continuously convey automation confidence to the operator. 

Results for a peg-in-hole application show that the sliding autonomy shared control approach increases error tolerance by at least $5\times$ over pure feed-forward automation.  The cost paid in completion time is $28$\% ($\sim$ $1.2$~s) of the mean completion time exhibited by the autonomous method.

Future work should consider estimating uncertainties online such that the system can achieve rapid completion time when uncertainty is low, but incorporate human intervention and maintain desired success rates when uncertainty is high.

\section{Acknowledgments}
This work was supported by Boeing Research and Technology (BR\&T) at the Boeing Advanced Research Center (BARC), University of Washington in Seattle.

\bibliographystyle{IEEEtran}
\bibliography{IEEEabrv,hmi}

\end{document}